\documentclass[sigconf]{acmart}

\usepackage[plain]{algorithm}
\usepackage{caption}
\usepackage{amsmath,amssymb,amsfonts}
\usepackage{graphicx}
\usepackage{textcomp}
\usepackage{xcolor}

\usepackage{tabularx}
\usepackage{float}
\usepackage{booktabs}

\usepackage{multirow}

\usepackage{subcaption}

\usepackage{listings,lstautogobble}
\usepackage[scaled=0.9]{DejaVuSansMono}

\lstdefinestyle{code}{
  frame=tb,
  numbers=left,
  numberstyle=\scriptsize\textsf,
  xleftmargin=2.2em,
  framexleftmargin=2.2em,
  basicstyle=\ttfamily\small,
  showstringspaces=false,
  commentstyle=\color{gray}
}

\newcommand{\key}[1]{\texttt{#1}}

\renewcommand\footnotetextcopyrightpermission[1]{} 
\begin{document}

%


\title{Solving All-Pairs Shortest-Paths Problem in Large Graphs\\ Using Apache Spark}

%
\author{Frank Schoeneman}
\affiliation{%
  \institution{Department of Computer Science \& Engineering\\University at Buffalo}
  \city{Buffalo}
  \state{New York}
}
\email{fvschoen@buffalo.edu}

\author{Jaroslaw Zola}
\affiliation{%
  \institution{Department of Computer Science \& Engineering\\Department of Biomedical Informatics\\University at Buffalo}
  \city{Buffalo}
  \state{New York}}
\email{jzola@buffalo.edu}

%
\begin{abstract}
Algorithms for computing All-Pairs Shortest-Paths (APSP) are critical building blocks underlying many practical applications. The standard sequential algorithms, such as Floyd-Warshall and Johnson, quickly become infeasible for large input graphs, necessitating parallel approaches. In this work, we propose, implement and thoroughly analyse different strategies for APSP on distributed memory clusters with Apache Spark. Our solvers are designed for large undirected weighted graphs, and differ in complexity and degree of reliance on techniques outside of pure Spark API. We demonstrate that the best performing solver is able to handle APSP problems with over 200,000 vertices on a 1024-core cluster. However, it requires auxiliary shared persistent storage to compensate for missing Spark functionality.
\end{abstract}

%
%



%

%
\maketitle

\section{Introduction}

All-Pairs Shortest-Paths is a classic problem in graph theory. The problem is both compute and memory intensive, owing to $O(n^3)$ computational and $O(n^2)$ space complexity for graphs with $n$ nodes. Over the years, many parallel APSP solvers have been proposed, including for distributed memory clusters, multi-core processors, and accelerators. However, to the best of our knowledge, currently no dedicated APSP approach is available for the Apache Spark model (see Section~\ref{sec:related}). This is somewhat surprising: many problems in Big Data analytics and Machine Learning (ML) -- domains in which Spark holds dominating position -- involve APSP as a critical kernel. For example, shortest paths in a neighborhood graph over high-dimensional points are known to be very robust approximation of geodesic distances on the underlying manifold~\cite{Bernstein2000}. Spectral dimensionality reduction methods, such as Multidimensional Scaling or Isomap~\cite{Tenenbaum2000}, directly involve APSP solver in their workflows. The same holds true for many other techniques, like networks classification~\cite{Borgwardt2005} or information retrieval~\cite{Quirin2008}. In all these applications, data sets with hundreds of thousands and even millions of points/vertices are not uncommon~\cite{Talwalkar2008,Schoeneman2018}. At the same time Spark is known and praised for its focus on programmer productivity and ease-of-use, including equally convenient deployments in HPC centers as in computational clouds. This makes Spark the frequent platform of choice for non-experts in parallel computing, who want to combine pure data analytics with processing of large graph instances.

In this paper, we ask the question of how efficient and scalable of a parallel APSP solver can we achieve with Apache Spark. We propose, implement and investigate four different APSP Spark-based solvers, which vary in how far they depart from what would be a {\it pure} Spark implementation. To facilitate the design, we identify and describe basic functional building blocks that when combined with Spark transformations and actions deliver a complete APSP solver in just a few lines of code, promoting programmer productivity. Then, we perform extensive experimental analysis of the solvers to assess efficiency and scalability of the resulting implementations on problems with up to $n=262144$ vertices using a 1,024-core cluster, and we demonstrate how proper choice of data partitioner and level of decomposition of graph adjacency matrix can be used to tune Spark performance. Our results show that using a {\it pure} Spark implementation, that does not involve auxiliary mechanisms and is fault-tolerant, is impractical for large problems due to the high cost of data shuffling. However, by leveraging collect and broadcast operations performed via auxiliary storage we are able to push the size of the problems we can solve, achieving good scalability.

We also contrast performance of Apache Spark with naive and highly optimized APSP implementations in MPI. Here our results show that although Spark is certainly able to handle large APSP problems, and can compete with naive MPI-based approach, it is clearly outperformed by MPI solvers with comparable level of optimization. This confirms a common wisdom that end-users should be aware of the trade off between programmer's productivity (and the ease of use), and the achievable~scalability.



The paper is organized following the common convention. In~Section~\ref{sec:related}, we briefly review related parallel APSP solutions. In~Section~\ref{sec:prelim}, we state basic assumptions and definitions we use in the work. In~Section~\ref{sec:methods}, we propose different strategies to implement APSP with Apache Spark and discuss their pros and cons taking into account implementation complexity and workarounds to missing Spark functionality. We describe detailed experimental analysis of the methods in Section~\ref{sec:results}, and conclude the paper with brief discussion and takeaway points in Section~\ref{sec:conclusion}.

\section{Related Work}\label{sec:related}

APSP remains a very actively researched problem in parallel computing, and hence we necessarily limit our review to projects that directly relate to~our~work.

In~\cite{Solomonik2013}, Solomonik et al. propose recursive 2D block-cyclic algorithm that achieves lower-bound on communication latency and bandwidth, which they next extend to 2.5D communication avoiding formulation. On a machine with 24,576 cores, the algorithms maintain strong scaling for problems with $n=32768$ nodes, and weak scaling for up to $n=131072$ nodes. The paper is also noteworthy for its extensive review of distributed memory APSP solvers. The advantage of the recursive formulation is induced data locality, which directly contributes to improved performance~\cite{Solomonik2013}. However, in Spark, the concept of data locality is much weaker, since Spark's runtime system has a significant freedom in scheduling where to materialize or move data for computations. Even though the programmer has some control over the data placement (e.g., via partitioning functions), they do not have direct control over where computations will be executed. In this work, we propose a solution akin to the iterative formulation of the 2D block-cyclic algorithm, which can be found in~\cite{Venkataraman2003}, and is one of several methods tracing their origin to transitive closure~\cite{Ullman1991}.

In~\cite{Katz2008}, Katz et al. give a solution of the transitive closure problem, and apply it to solving APSP on a GPU. They demonstrate 2-4$\times$ speedup over a highly tuned CPU implementation, and report problems for up to $n < 10000$. Another efficient GPU-based solution is given by Djidjev et al. in~\cite{Djidjev2014}. However, the method applies only to planar graphs. Djidjev solves APSP on sparse graphs with up to one million nodes, using a combination of multi-core CPUs and two~GPUs. While we could combine GPU acceleration with Spark, e.g., by using Python and Numba Just-in-Time (JIT) compiler~\cite{Lam2015}, it is not entirely clear how advantageous such approach would be.

APSP is one of several graph primitives that can be directly posed as a linear algebra problem, and solved using matrix operations over the semi-ring $(\min, +)$~\cite{Kepner2011}, e.g., by employing GraphBLAS~\cite{Kepner2016}. Since asymptotically efficient matrix multiplication algorithms (e.g., Strassen) are not applicable to the general APSP under min-plus matrix multiplication, algorithms to improve on $O(n^{3})$ complexity have been proposed for cases when the graph of interest is directed, undirected and unweighted, or undirected with integer weights from a finite set~\cite{Seidel1995,Shoshan1999,Zwick1998}. In our work, we directly exploit the semi-ring formulation as one possible implementation. Additionally, as in the case of GPU acceleration, we could combine multi-core GraphBLAS implementation with Spark backbone (see Section~\ref{sec:methods}).

Apache Spark frameworks, such as GraphX~\cite{Xin2013} (no longer maintained) and GraphFrames~\cite{Dave2016}, offer multi-source shortest-paths algorithms. These algorithms are simple extensions of the single source shortest paths solver in the Pregel/BSP model~\cite{Malewicz2010}, and are not designed with APSP in mind. Our proposed solutions break away from the Pregel model, in favour of 2D blocked decompositions, because in the initial tests GraphX was unable to handle any reasonable problem size, prompting us to investigate alternative approaches.

\section{Preliminaries}\label{sec:prelim}


We focus on computing length of all pairs shortest paths (i.e., no paths themselves) in undirected weighted graphs with no negative cycles. Let $G=(V,E,w)$ be such an undirected graph, where $V$ is a set of $n$ vertices, $E$ is a set of edges, and $w:E \rightarrow \mathbb{R}$ determines edge weights. At this point we want to stress that we make no assumption regarding graph sparsity, structure (e.g., planar), or weights distribution. This is because many Spark-based data analytics pipelines that involve APSP are general purpose, and have to be responsive to graphs with varying properties.

In general, and in ML or Big Data analytics in particular, each vertex $v \in V$ might be a complex object with non-trivial memory layout. However, we assume that some initial pre-processing of the input graph has been performed, and each vertex is uniquely identified by an integer index. This is inline with the common use patterns, where APSP is invoked as a computational building block within a larger framework. The standard approach to handle APSP in such cases is to use a variant of classic Floyd-Warshall~\cite{Cormen2009} or Johnson algorithms~\cite{Cormen2009}, with complexity $O(|V|^3)$ and $O(|V||E| + |V|^2\log(|V|))$, respectively. The Johnson algorithm offers better asymptotic behaviour for reasonably sparse graphs, but typically Floyd-Warshall derivatives outperform it as they allow for better computational density (see also~\cite{Solomonik2013}). When using Floyd-Warshall and related algorithms, we will represent adjacency matrix of $G$ by $A$, where $A_{ij} = A_{ji} = w_{ij}$ stores the weight of edge between vertices with indices $i$~and~$j$. Furthermore, we are going to use the following notation. We will write $A_{IJ}$ to describe block $(I,J)$ of matrix $A$ (size of the block and indexing order will be clear from the context), and $A_{I\cdot}$ and $A_{\cdot J}$ to describe row-block $I$ and column-block~$J$, respectively. Similarly, we will write $A_{i\cdot}$ and $A_{\cdot j}$ to describe row $i$ and column $j$ of $A$, and $A_{Ik}$ and $A_{kJ}$ to describe column $k$ in row-block $I$ and row $k$ in column-block $J$.

To maintain all data elements in our proposed solutions, e.g., matrix $A$, we will primarily depend on Spark's Resilient Distributed Datasets (RDDs)~\cite{Zaharia2012}. For the sake of completeness: an RDD is a fault-tolerant abstraction over distributed memory, tailored for in-memory iterative algorithms. RDDs are non-mutable, and are lazily evaluated and transparently materialized either in main memory or persistent storage. Unless explicitly specified by a programmer, their partitioning is automatically handled by the Spark runtime using either keys range or hash-based partitioning schemes. From the computational point of view, objects within an RDD can be processed in parallel by Spark executors using a set of transformations that yield new RDDs. A Spark driver node is responsible for managing RDD lineage, and orchestrating the Spark application execution.

When implementing APSP solvers, whenever possible we will depend solely on Apache Spark API, benefiting from the implicit fault-tolerance ensured by the platform. Our focus will be on programmer-perceived productivity (e.g., minimizing boilerplate code, using convenient high-level abstractions, etc.). We will refer to such implementations as {\it pure}. To improve performance, in some implementations we will depend on workarounds to missing Spark functionality (e.g., implementing point-to-point data exchange through a shared file system). When such implementations will not be fault-tolerant (e.g., failed tasks depending on data in a shared file system are not guaranteed to be able to access that data when rescheduled), we will refer to them as {\it impure}.

\section{Proposed APSP Spark Solvers}\label{sec:methods}

We propose and investigate four different APSP approaches that vary in the implementation complexity (and as we shall see, in performance). In all approaches, we 2D decompose adjacency matrix $A$ into $q \times q$ blocks, where $q = \lceil\frac{n}{b}\rceil$, and $b$ is a user-provided (or auto-tuned) decomposition parameter. We store the resulting matrix blocks as key-value tuples $((I,J),A_{IJ})$ in Spark RDDs, where tuple $(I,J)$ is a key we will use to reference the corresponding data block. We will store each block $A_{IJ}$ as a dense matrix. This is because, in practical cases, $A$ very quickly becomes dense matrix, and hence potential savings from using a sparse format in early iterations are negligible. 
The matrix decomposition will be different from RDD partitioning, in that a single RDD partition will be maintaining multiple matrix blocks. We will use different ways of assigning RDD partitions to executors (see Section~\ref{sec:results}), which we can think of as over-decomposition of $A$. Moreover, even though we will be discussing our approaches as if the entire matrix $A$ was stored in an RDD (to simplify presentation), in the actual implementation (and hence experiments) we exploit symmetricity of $A$ and store only its upper-triangular part. The remaining blocks are generated on-demand by transposition (with no measurable overheads). Hence, the executor responsible for the processing of block $A_{IJ}$ is also responsible for the processing of block $A_{JI}$. This enables us to reduce the total amount of data maintained by the RDD, while increasing computational costs of processing~tasks. We note that by disregarding symmetricity of $A$, our algorithms can be directly adopted for cases where $G$ is a directed graph.

To implement our proposed algorithms we use pySpark. The choice is motivated by practical considerations. First, the use of Python makes it convenient and efficient to store and communicate data blocks maintaining C/C++ compatible row-major matrix representation (e.g., via NumPy arrays). This in turn enables us to easily offload computationally intensive operations to a bare-metal runtime, either using direct C/C++ bindings (e.g., CPython, Boost.Python~\cite{Abrahams2003}, implicitly SciPy~\cite{Jones2014}), or through the excellent Numba JIT compiler~\cite{Lam2015} (see next subsection). Second, serialization and deserialization of Python objects, which plays a role in our implementations (e.g., when exchanging data via persistent storage), is supported out-of-the-box, further contributing to the ease of implementation and programmer's productivity. We recognize the fact that pySpark tends to incur overheads due to object conversions between Python and Spark's native Scala/JVM runtime~\cite{Xin2015}. However, as we mentioned earlier, in this work we opt for productivity and compactness of the Python code while leveraging bare-metal execution whenever possible.

\subsection{APSP Functional Building Blocks}

To separate computational parts of an APSP solver from the pure Spark mechanics, we first identify functional blocks that will be shared between different implementations and will facilitate algorithms design. These building blocks can be thought of as fundamental operations performed by APSP algorithms, and ideally should be highly optimized and delegated to bare-metal hardware for execution. Such semantic separation is advantageous not only from the programmer's point of view (delivering reusable and compact abstractions), but also from the algorithms design and specification point of view (e.g., simplifying reasoning about correctness, computational complexity and space complexity).

We present each identified element as a function (we use term function somewhat liberally, since in our case a function may return/yield multiple elements for the same argument). The functions will be passed to Spark transformations to act directly on matrix blocks stored by an RDD, in which case their first argument, omitted from invocation, will always be a single record stored in the RDD for which transformation is invoked. Table~\ref{tab:elements} summarizes all the functions (including their syntax and short description). How they are used will become clear in the following~subsections. We note that our functional blocks are different from the concept of graph building blocks~\cite{Kepner2016}, that are tailored to express much broader spectrum of graph algorithms (compared to our explicit focus on APSP).

\begin{table}[ht]
\centering
\caption{Summary of our functional elements.}\label{tab:elements}
\renewcommand{\arraystretch}{1.75}
\begin{tabular}{p{0.9\columnwidth}}
\hline
$\key{InColumn}[((I,J),A_{IJ}), x]: \key{return}~J = x;$\newline
Predicate testing whether given block is in column-block $x$. \\
\hline
$\key{OnDiagonal}[((I,J),A_{IJ}), x]: \key{return}~(I = J)~\&~(I = x);$\newline
Predicate testing if given block is $x$-th block on diagonal. \\
\hline
$\key{ExtractCol}[((I,J),A_{IJ}), k]: \key{return}~(I, A_{I(J\cdot b + k)});$\newline
Return a new record with $k$-th column of given block (the column is stored as a vector). Within a block rows and columns are zero-indexed. \\
\hline
$\key{CopyDiag}[((I,I),A_{I,I})]: \key{forall}~J~\key{yield}~((I,J),A_{II});$\newline
Create $(q - 1)$ copies of a diagonal block, each identified by different column-block index. \\
\hline
$\key{CopyCol}[((I,J),A_{IJ})]:\newline
~~~\key{forall}~K~\key{yield}~((I,K),A_{IJ})$ and $~\key{yield}~((K,I),A_{IJ}^{T});$\newline
Create $(q-1)$ copies of a given block, each identified by different column-block index, and $(q-1)$ copies of its transpose, each identified by different row-block index. \\
\hline
$\key{MatMin}[((I,J),A_{IJ}),B]: \key{return}~((I,J),\min(A_{IJ},B));$\newline
Return element-wise minimum between block $A_{IJ}$ and $B$. \\
\hline
$\key{MatProd}[((I,J),A_{IJ}), B]: \key{return}~((I,J),A_{IJ} \otimes B);$\newline
Return min-plus product between block $A_{IJ}$ and $B$. \\
\hline
$\key{MinPlus}[((I,J),A_{IJ}), B]:\newline
~~~\key{return}~((I,J),(\min((A_{IJ} \otimes B), B);$\newline
Returns $\key{MatProd}$ followed by $\key{MatMin}$ for block $A_{IJ}$ and $B$. \\
\hline
$\key{FloydWarshallUpdate}[((I,J),A_{IJ}),B_{Ik},B_{Jk}]:\newline
~~~C=B_{Ik}\cdot 1^{T} + 1\cdot B_{Jk}^{T};~\key{return}~\key{MatMin}(((I,J),A_{IJ}),C);$ \\
\hline
$\key{FloydWarshall}[((I,I),A_{I,I})]: \key{return}~((I,I),A'_{I,I});$\newline
Execute Floyd-Warshall, or any other APSP solver, over diagonal block~$A_{I,I}$, and return the resulting distance matrix $((I,I), A'_{I,I})$. \\
\hline
$\key{ListAppend}[List,((I,J),A_{IJ})]:\newline
~~~\key{return}~List.\key{append}(((I,J),A_{IJ}));$\newline
Append block $A_{IJ}$ to list of blocks $List$. This function is a simplified representation of the actual Spark combiner.\\
\hline
$\key{ListUnpack}[((I,J),List)]:\newline
~~~(A_{IJ}, B_1, B_2) = List;\newline
~~~\key{if undefined}(B_2): \key{return}~(((I,J),A_{IJ}), A_{IJ} \otimes B_1);\newline
~~~\key{else}: \key{return}~(((I,J),A_{IJ}), B_1 \otimes B_2);$\newline
Process list of blocks to return block that will be second argument of min-plus product with $A_{IJ}$. \\
\hline
\end{tabular}

\smallskip
\small{We use \texttt{yield} to indicate that multiple elements are returned.}
\end{table}

\subsection{Repeated Squaring}

In the first approach, we exploit APSP min-plus product formulation, and we essentially compute $A^n$ using the classic repeated squaring method. While asymptotically this method is clearly inefficient, it is very fast to implement and highlights programmer productivity when using Spark. It also introduces how functional blocks are combined with Spark API.

Given an RDD with block decomposed $A$, repeated squaring becomes a sequence of three steps over the RDD: \key{cartesian} followed by \key{filter} to group blocks that should be multiplied in the current iteration, \key{map} applying min-plus product (function \key{MatProd}), and finally \key{reduceByKey} with function \key{MatMin} to finalize the product. All blocks of $A$ can be persisted in the total main memory of the executing cluster, hence directly leveraging support for iterative algorithms in Spark (this will be the case for the remaining algorithms as well). However, the problem with this approach is reliance on $\key{cartesian}$ that involves extensive all-to-all data shuffle. In our tests, we found that $\key{cartesian}$ was easily stalling even on small problems. Hence, to bypass this bottleneck of {\it pure} implementation, we replaced \key{cartesian} with iteration over the column-blocks of $A$, effectively rewriting matrix-matrix product into a series of matrix-vector products. This allows us to reduce data movement, at the cost of increased Spark overheads (like scheduling delays or tasks deserialization costs).

\begin{algorithm}[t]
\centering
\caption{Repeated squaring with column-blocks.}\label{alg:repeatedsqr}
\begin{lstlisting}[style=code,language=Python]
for i in range(0, log(n)):
  for J in range(0, q):
    col = A.filter(InColumn(J)).collect()
    for block in col: block.tofile()
    T[J] = A.map(MatProd).reduceByKey(MatMin)
  A = sc.union(T)
\end{lstlisting}
\end{algorithm}

The core of the resulting implementation is outlined in Algorithm~\ref{alg:repeatedsqr}. In line 3, we identify blocks of column-block $J$ to multiply, and group them on the Spark driver node, which next distributes the entire column to executors (line 4). Note that we do not broadcast the column, but rather store its blocks in a shared file system available to driver and executor nodes  (e.g., HDFS, GPFS, etc.). As a result, the appropriate blocks can be selected and used by executors only when needed (most of the executors will not be using all blocks). In line 5, we perform the actual matrix-vector product. Here, \key{MatProd} takes its first argument, block $A_{IK}$, directly from the RDD, and the second argument is the $K$-th block of the column-block~$J$ deserialized from the shared secondary storage. Both $\key{MatProd}$ and $\key{MatMin}$ are delegated for bare-metal execution using Numba and NumPy, respectively. The results of $q$ individual matrix-vector products are brought together via \key{union} (line 6) to form the RDD ready for the next iteration of the algorithm.

 Our implementation of repeated squaring does not require any particular attention to, e.g., when RDDs are materialized, delegating the entire execution to Spark. However, utilizing persistent storage for broadcast introduces side-effects, which go against Spark's fault-tolerance mechanisms, making the implementation {\it impure}. We note that one could consider using shared file system directly at the filtering stage, eliminating the need for \key{collect} on the driver. However, this approach is problematic considering the Spark scheduler, e.g., files written in one RDD may not be materialized on time to be used by a subsequent RDD.

The computational cost of repeated squaring is bounded by $\frac{n}{b}\log(n)$ iterations, where each iteration involves \key{map} transformation with $O(b^3)$ operation, followed by reduction.

\subsection{2D Floyd-Warshall}

In our second approach, we adopt the textbook parallel Floyd-Warshall algorithm based on 2D block decomposition~\cite{Grama2003}. In this approach, parallelism is due to the two innermost loops of Floyd-Warshall, which can be partitioned once row/column indexed by the outermost loop is block-wise broadcast, and locally available to the processors. Since Spark provides \key{collect} and \key{broadcast} that combined can emulate standard broadcast initiated at any processor, we can easily express the entire algorithm in Spark. 

\begin{algorithm}[t]
\centering
\caption{Floyd-Warshall with 2D decomposition.}\label{alg:fw}
\begin{lstlisting}[style=code,language=Python]
for k in range(0, n):
  K = k / b
  kloc = k % b

  D = A.filter(InColumn(K)).map(ExtractCol(kloc)) \
       .collect()

  colk = sc.broadcast(D)

  A = A.map(FloydWarshallUpdate)
\end{lstlisting}
\end{algorithm}

Our Spark implementation is outlined in Algorithm~\ref{alg:fw}. The algorithm proceeds in $n$ steps. In iteration $k$, we first identify column blocks storing column $k$, and then we extract the actual column using function \key{ExtractCol} (lines 2-5). The column is aggregated on the driver node (line 6) and broadcast to all executors (line 8). Because the memory footprint of a column is very small, the operation can be easily performed without the need for persistent storage. This step is an explicit synchronization point. Once column $k$ is available to every executor, we proceed with the update phase of Floyd-Warshall algorithm, as implemented in \key{FloydWarshallUpdate} function. In this case, the first argument is block~$A_{IJ}$ taken directly from the RDD, and the two remaining arguments are vectors extracted from the column $k$ (exploiting the fact that $A$ is symmetric). As in earlier algorithm, the actual computations are delegated for bare-metal execution using Numba JIT compilation.

The 2D Floyd-Warshall is interesting in the sense that it is amenable to {\it pure} Spark implementation, supporting fault-tolerance, and involving no side-effects. Moreover, it does not require any so called Spark wide transformations, which trigger data shuffling and movement. Consequently, it plays into Spark strengths. However, its computational cost is bounded by $n$ iterations, where each iteration involves building blocks with cost $O(b^2)$. Since synchronization overheads in Spark are significant, we anticipate poor scalability.

\subsection{Blocked In-Memory}

\begin{figure}
    \centering
    \includegraphics[scale=0.9]{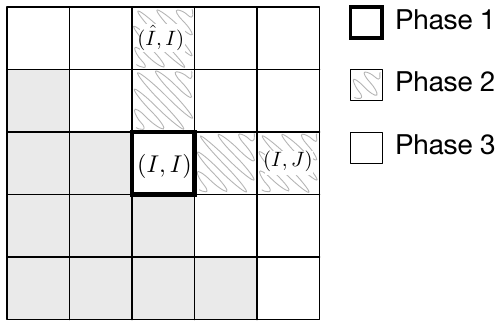}
    \caption{Processing phases in the blocked Floyd-Warshall algorithm. Diagonal blocks form critical path.}
    \label{fig:phases}
\end{figure}

In the third approach, we use ideas from the blocked APSP from Venkataraman et al.~\cite{Venkataraman2003}. Originally, the algorithm was designed to improve cache utilization of the standard Floyd-Warshall. It is an iterative algorithm, where each iteration consist of three phases, as depicted in Figure~\ref{fig:phases}. In Phase~1, diagonal block $A_{II}$ is processed using any efficient APSP solver (in our case sequential Floyd-Warshall). In Phase~2, the result is passed to update shortest paths in all blocks of block row $A_{I\cdot}$ and block column $A_{\cdot I}$. Then, in Phase~3, iteration is completed by performing update on the remaining blocks.

\begin{algorithm}[t]
\centering
\caption{Blocked In-Memory.}\label{alg:blockedmem}
\begin{lstlisting}[style=code,language=Python]
for i in range(0, q):
  diag = A.filter(OnDiagonal(i))
  d = diag.map(FloydWarshall).flatMap(CopyDiag) \ 
          .partitionBy(partitioner, p)

  rowcol = A.filter(InColumn(i))
  d = sc.union([d, rowcol]) \
        .combineByKey(ListAppend).map(ListUnpack) \ 
        .map(MatMin).flatMap(CopyCol) \ 
        .partitionBy(partitioner, p)

  A = A.filter(not InColumn(i)).union(d) \ 
       .combineByKey(ListAppend) \ 
       .map(ListUnpack).map(MatMin) \ 
       .partitionBy(partitioner, p)
\end{lstlisting}
\end{algorithm}

Our take on the algorithm, which we call Blocked In-Memory, is outlined in Algorithm~\ref{alg:blockedmem}. In $q$ iterations, we start by processing a diagonal block (lines~2-4) using \key{FloydWarshall} function executed on bare-metal via a call to an optimized SciPy routine. The diagonal block is extracted from RDD via filtering, and once processed we explicitly create its $q - 1$ copies (function \key{CopyDiag}) that are distributed to executors via a custom partitioner (line~4). Here our goal is to place these copies in the RDD partitions that store blocks from the corresponding column/row block (extracted in line~6), thus maximizing data locality and mitigating partition blowup due to RDDs merging in line~7 (we provide details of partitioning schemes in the next section). We can think about this step as using data shuffling to simulate general broadcast. This workaround is necessary because Spark API does not expose broadcast to executors. Overall, this phase of the algorithm stresses Spark's ability to optimize data distribution and corresponding tasks scheduling. To complete pairing of diagonal and column/row blocks we use a combination of \key{combineByKey} and \key{map} (line 8), by first aggregating a list of blocks to pair (\key{ListAppend}), and then enumerating the actual block pairs (\key{ListUnpack}). The resulting pairs are processed to compute the update (line~9), and as previously we use block copying and custom partitioning to bring together blocks that will be operated on in Phase 3 (lines 9-10). To finalize iteration, we implement Phase 3 computations (lines 12-15) following exactly the same pattern (with blocks pairing) as in case of Phase 2.

The above implementation of our blocked algorithm depends entirely on fault-tolerant Spark functionality, and hence we consider it {\it pure}. While it operates in $q$ iterations, it is data intensive due to data copying and shuffling involved. However, the method should be able to leverage significant data locality, hence it is a good test for the Spark runtime system.

The computational cost of the algorithm is bounded by $O(\frac{n}{b})$ iterations, with each iteration bounded by $O(b^3)$. While asymptotically this is similar to 2D Floyd-Warshall, the blocked algorithm allows us to control (through the parameter $b$) the trade-off between the the cost of single iteration and the number of iterations (which is not the case in the previous methods).

\subsection{Blocked Collect/Broadcast}

Our last solver, given in Algorithm~\ref{alg:blockedbcst}, is a redesign of the Blocked In-Memory APSP to bypass explicit data shuffling. Specifically, instead of pairing a diagonal block with \mbox{column/row} blocks, we bring it to the driver node via \key{collect}, which then redistributes it to the executors via shared persistent storage (lines 2-3). Hence, we eliminate costly all-to-all data exchange with a communication through the Spark driver node (which should be faster at the expense of using persistent storage). We note that here, similar to the repeated squaring algorithm, we do not use \key{broadcast} because in pySpark each task created by an executor maintains its local copy of the broadcast variables. This usually means exceeding the executor's memory, since the number of running tasks may be the same as the number of cores in a node running the executor. To realize Phase~2, we apply \key{MinPlus} function on the column blocks such that the first argument comes from the RDD, and the second argument is taken from Spark storage (line~5). As in earlier phase, we aggregate the entire column on the driver node and then redistribute it via shared persistent storage~(lines 6-7). Finally, in line~9, we perform Phase~3 computations updating the remaining blocks of $A$ with the column blocks from persistent storage. All updated blocks are brought into a single RDD via \key{union}, and repartitioned to match intended partitioning of $A$ (lines~11-12).

\begin{algorithm}[t]
\centering
\caption{Blocked Collect/Broadcast.}\label{alg:blockedbcst}
\begin{lstlisting}[style=code,language=Python]
for i in range(0, q):
  diag = A.filter(OnDiagonal(i)).map(FloydWarshall)
  diag.collect().tofile()
  
  rowcol = A.filter(InColumn(i)).map(MinPlus)
  rowcol_coll = rowcol.collect()
  for b in rowcol_coll: b.tofile()
  
  offcol = A.filter(not InColumn(i)).map(MinPlus)

  A = sc.union([diag, rowcol, offcol]) \ 
        .partitionBy(partitioner, p)
\end{lstlisting}
\end{algorithm}

The collect and broadcast implementation of the blocked algorithm involves secondary storage to handle communication, and hence we consider it {\it impure}. Except of custom partitioning of matrix $A$, we do not repartition RDD that stores column/row blocks (line 5). Instead we depend on the partitioning scheme of $A$ from which we extract blocks. This means that Spark runtime will most likely trigger data movement to utilize all executors.

Asymptotically, the cost of the algorithm is the same as Blocked In-Memory. However, in single iteration we substitute data shuffling with writing to auxiliary storage.

\section{Experimental Analysis}\label{sec:results}


To benchmark all four algorithms and their implementations, we performed a set of experiments on a standalone Apache Spark cluster with 32 nodes and GbE interconnect. Each node in the cluster is equipped with two 16-core Intel Skylake (Intel Xeon Gold 6130 2.10GHz) processors with 32KB L1 and 1024KB L2 cache, and 192GB of RAM (thus the cluster provides total of 1,024 cores and 6TB of RAM). Moreover, each node has standard SSD drive available, which Spark uses for local data staging (available local storage is 1TB). We note that the use of SSDs is essential for Spark performance, since all data movement in Spark involves staging in the local storage. The local storage is complemented by shared GPFS storage that executors can use for additional communication (e.g., as in case of Blocked Collect/Broadcast algorithm).

In all tests, the Spark driver was run with 180GB of memory, to allow efficient management of complex RDD lineages. For the same reason, we run the driver on a dedicated node separately and in addition to Spark executors. We allocated one executor per node using the default configuration for the number of cores, i.e., each executor was using all available cores in a node. All executors were configured to use 180GB out of the available 192GB, with the remaining memory available to the operating system and Python interpreter. While large memory available to Spark runtime could potentially lead to overheads in garbage collection, we did not observe this in practice. We note that we tested different Spark runtime configurations, including executor-to-core ratio, memory use fractions, etc.~\cite{SparkTuning}, without noticeable difference in~performance.

Our entire software is implemented in Apache Spark 2.2 pySpark, and Python 2.7. Compute intensive linear algebra operations and sequential Floyd-Warshall algorithm are offloaded to bare-metal via NumPy and SciPy that are configured to work with the Intel MKL~2017 BLAS library. Finally, we use Numba~0.35 for just-in-time~compilation whenever required. Our entire software is open source and available from \url{https://gitlab.com/SCoRe-Group/APSPark} together with all benchmark data (see below).

\subsection{Test Data}

All four APSP solvers are oblivious to the structure of input graph, and in our implementations we do not include optimizations to target any specific graph properties (recall that we represent the graph using dense matrices). Consequently, the scalability of every solver is a function of graph size expressed only by the number of vertices, $n$. Therefore our input data is simply a set of synthetic Erd\H{o}s-R\'enyi graphs. In all graphs, the probability of edge,~$p_e$, is set to $p_e = \frac{(1+\epsilon)\cdot \ln(n)}{n}$, with $\epsilon = 0.1$. The particular choice of $\epsilon$ is to make graphs generation fast, and we report it only for the sake of reproducibility. We note that taking $p_e$ larger to increase likelihood of connectivity in no way adversely affects our performance, as we disregard the cost of populating RDD that stores the adjacency matrix $A$, and each of our approaches scales only with $n$. In our tests, we consider graphs for $n$~up to 262,144.

At this point we wish to reiterate that the performance of our methods is {\bf in no way impacted} by any properties of the graph aside from the number of vertices $n$.

\subsection{Block Size and Sequential Components}\label{sec:seqcompperf}

In the first set of experiments, we setup a baseline for subsequent comparisons by analyzing performance of the key functional elements, \key{MatProd} combined with \key{MatMin} and \key{FloydWarshall}, depending on the block size $b$. These building blocks will be dispatched by Spark for sequential execution, and hence their performance will be critical to the overall performance. Moreover, in all our implementations, the block size directly affects level of available parallelism, and level of data movement. If the block size is too large, we may be spending too much time in a sequential execution or staging data blocks in the auxiliary storage, and we may not be able to leverage all executors. On the other hand, if block size is too small, overheads due to tasks scheduling and data shuffling, or creating many small files in the auxiliary storage, may dominate the execution. Hence, the block size should be selected carefully.

\begin{figure}[b]
  \centering
  \includegraphics{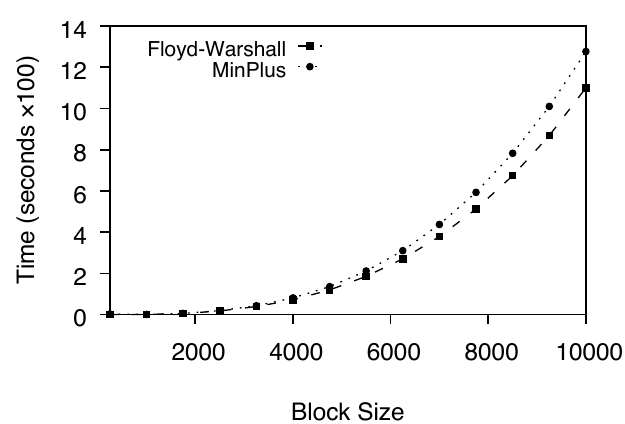}
  \caption{The effect of block size on the execution time of sequential operations $\key{MatProd}$ combined with $\key{MatMin}$, and $\key{FloydWarshall}$.}\label{fig:mpbvstime}
\end{figure}

\begin{figure}[ht]
  \centering
  \includegraphics{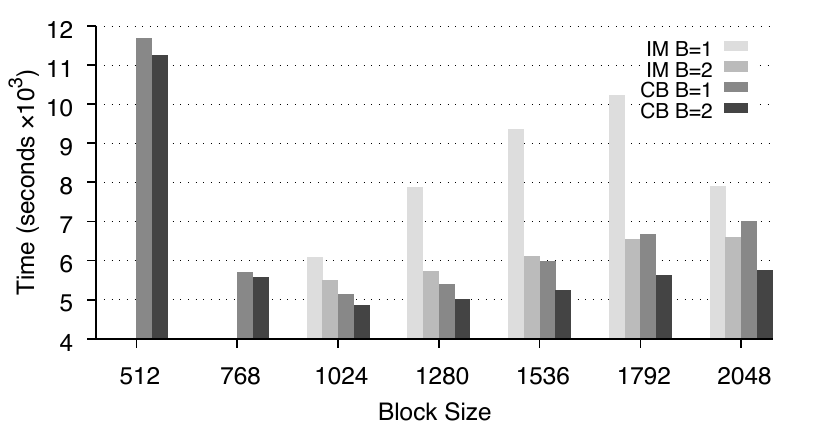}
  \includegraphics{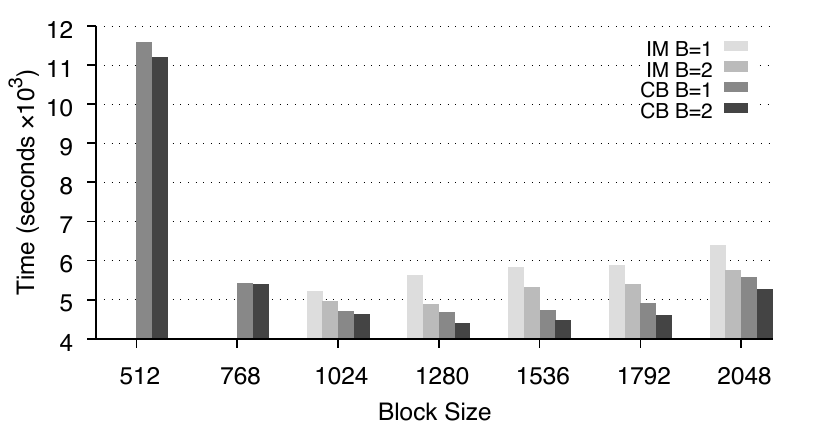}
  \includegraphics{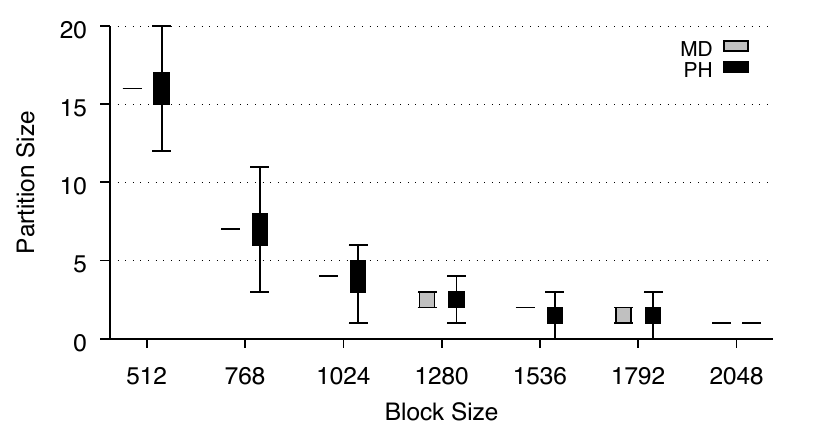}
  \caption{Example effect of block size on the execution time of Blocked In-Memory~(IM) and Blocked Collect/Broadcast~(CB), depending on the choice of block size, partitioner and partition size. Top: default Spark partitioner. Middle: our multi-diagonal partitioner. Bottom: distribution of RDD partition sizes incurred by both partitioners when $B=2$. Problem size $n=131072$ run on $p=1024$~cores. $B$ indicates the number of RDD partitions per core.}\label{fig:blockandpart}
\end{figure}

To test the sequential components, we call the corresponding functions directly from Python, the same way as they would be invoked~by~Spark, i.e., \key{FloydWarshall} via SciPy with Intel MKL, and \key{MatProd} and \key{MatMin} via Numba. To test the actual solvers, we use our pySpark implementations. In all tests, we look at the observed execution times. Results of these experiments are reported in Figures~\ref{fig:mpbvstime}~and~\ref{fig:blockandpart}.

Figure~\ref{fig:mpbvstime} shows how block size, $b$, affects the time of processing a single adjacency matrix block. As expected, the runtime increases roughly as $O(b^3)$, in line with the asymptotic behavior of the underlying algorithms. For $b$ up to approximately 3,000, sequential operations can be executed very quickly, primarily because adjacency matrix fits in cache memory. Given the available Intel Skylake CPU cache, the approximate size for processing completely in L3 cache is around~$b=1810$. Once $b$ is above that threshold, runtime starts to grow rapidly, going into minutes.

Figure~\ref{fig:blockandpart} highlights the impact of partitioning granularity on the total execution time. Here we focus on Blocked In-Memory~(IM) and Blocked Collect/Broadcast~(CB) methods, as these are the best performing methods (see next subsection).

Irrespective of the choice of RDD partitioner (we discuss partitioners later), the runtime of both methods first decreases, and then grows with the growing block size. When the block size is too small ($b < 1024$), the Blocked In-Memory solver fails. This is because the volume of the data that has to be staged in local SSDs due to shuffling exceeds the storage capacity of our cluster (in our case, each node is equipped with 1TB of local storage). Specifically, data repartitioning (call to \key{partitionBy} in Algorithm~\ref{alg:blockedmem}, line~15) triggers data shuffling in each iteration. Because shuffled blocks are spilled to the local storage and preserved for fault tolerance, the storage requirement grows linearly with the number of iterations. At the same time, eliminating the call to \key{partitionBy} is not an option: without repartitioning step the number of partitions in the RDDs created via \key{union} (Algorithm~\ref{alg:blockedmem}, line 12) would quickly explode since in Spark each component RDD preserves its partitioning when in \key{union}. The excessive number of partitions would degrade performance of the \key{combineByKey} step, and would add significant scheduling overheads. We note that this issue does not affect Blocked Collect/Broadcast, because the volume of the data we manage is smaller (recall that here we avoid creating copies of the data blocks), and repartitioning is thus less aggressive.

The results above confirm the impact of the block size on the performance of our APSP solvers. They suggest also that accelerating single block computations, e.g., by delegating them to GPGPU, could further improve performance of the solvers. Specifically, since with acceleration we should be able to process larger blocks in the same time limit, we could reduce the number of iterations required by the solver without significantly affecting the cost of individual~iterations.

\subsection{Tuning Performance of the Solvers}\label{ssec:partfunc}

In the next set of experiments, we take a closer look at the performance of our solvers in relation to RDD partitioning scheme that we use to distribute RDD with matrix $A$. In Figure~\ref{fig:blockandpart}, we can observe that both partitioning function as well as the number of RDD partitions per core (parameter $B$) affect the runtime, and hence tuning them may lead to better performance.

The efficiency of each of our APSP methods depends on data locality. Ideally, to reduce the number and frequency of Spark data shuffles, blocks that are paired for computation should be assigned to the same partition, and partitions should be evenly distributed between executors. We should keep in mind however, that even then there is no guarantee that the executor maintaining given partition will be responsible for its processing. Thus, to provide a room for load balancing, the ratio between the number of partitions and the number of available cores should be more than one (per Spark guidelines~\cite{SparkTuning}, the number of partitions is recommended to be $2\times$-$4\times$ the total number of cores, i.e., $2 \leq B \leq 4$).

\begin{table}[t]
\caption{The effect of block size on execution time.}\label{blockmethodstable0}
\centering
{\footnotesize
\begin{tabularx}{\columnwidth}{lXXXXX}
\toprule
\multicolumn{1}{c}{Method} & Partitioner & \multicolumn{2}{c}{Decomposition} & \multicolumn{2}{c}{Time}\\
~ & ~ & $b$ & Iterations & Single & Projected\\
\midrule
\multirow[t]{10}{*}{Repeated Squaring} & \multirow[t]{5}{*}{MD}& 256 & 18432 & 45s & \textbf{9d16h}\\
& & 512 & 9216 & 2m23s & 15d8h\\
& & 1024 & 4608 & 5m6s & 16d8h\\
& & 2048 & 2304 & 19m45s & 31d15h\\
& & 4096 & 1152 & 51m47s & 41d10h\\
\cmidrule{2-6}
& \multirow[t]{5}{*}{PH}& 256 & 18432 & 44s & \textbf{9d11h}\\
& & 512 & 9216 & 2m7s & 13d13h\\
& & 1024 & 4608 & 6m5s & 19d12h\\
& & 2048 & 2304 & 18m39s & 29d21h\\
& & 4096 & 1152 & 1h15m & 60d6h\\
\midrule
\multirow[t]{10}{*}{2D Floyd-Warshall} & \multirow[t]{5}{*}{MD}& 256 & 262144 & 21s & 64d11h\\
& & 512 & 262144 & 18s & 53d10h\\
& & 1024 & 262144 & 17s & \textbf{51d22h}\\
& & 2048 & 262144 & 18s & 55d7h\\
& & 4096 & 262144 & 20s & 61d9h\\
\cmidrule{2-6}
& \multirow[t]{5}{*}{PH}& 256 & 262144 & 21s & 65d8h\\
& & 512 & 262144 & 18s & 55d10h\\
& & 1024 & 262144 & 16s & \textbf{49d7h}\\
& & 2048 & 262144 & 20s & 60d3h\\
& & 4096 & 262144 & 19s & 56d9h\\
\midrule
\multirow[t]{10}{*}{Blocked-IM} & \multirow[t]{5}{*}{MD}& 256 & 1024 & 51s & 14h29m\\
& & 512 & 512 & 1m11s & 10h8m\\
& & 1024 & 256 & 1m55s & 8h12m\\
& & 2048 & 128 & 3m44s & 7h59m\\
& & 4096 & 64 & 7m21s & \textbf{7h51m}\\
\cmidrule{2-6}
& \multirow[t]{5}{*}{PH}& 256 & 1024 & 48s & 13h32m\\
& & 512 & 512 & 1m14s & 10h33m\\
& & 1024 & 256 & 2m12 & 9h23m\\
& & 2048 & 128 & 4m3s & \textbf{8h39m}\\
& & 4096 & 64 & 8m49s & 9h24m\\
\midrule
\multirow[t]{10}{*}{Blocked-CB} & \multirow[t]{5}{*}{MD}& 256 & 1024 & 48s & 13h35m\\
& & 512 & 512 & 1m1s & 8h40m\\
& & 1024 & 256 & 1h40m & 7h8m\\
& & 2048 & 128 & 3m18s & \textbf{7h4m}\\
& & 4096 & 64 & 8m23s & 8h57m\\
\cmidrule{2-6}
& \multirow[t]{5}{*}{PH}& 256 & 1024 & 46s & 13h12m\\
& & 512 & 512 & 1m3s & 9h4m\\
& & 1024 & 256 & 1m51s & \textbf{7h54m}\\
& & 2048 & 128 & 3m51s & 8h15m\\
& & 4096 & 64 & 9m23s & 10h2m\\
\bottomrule
\end{tabularx}}
\smallskip
{\scriptsize $n=262144, p=1024, B=2$}
\end{table}%

In our implementation we consider two partitioners. The first one is the default Spark partitioner, called Portable Hash (PH), which is available in pySpark as method $\key{portable\_hash}$. This is the partitioner that one would use {\it ad hoc}. The partitioner is expected to distribute RDD uniformly at random by computing a hash code for Python tuples $(I,J)$, which are used as keys in all our RDDs. The resulting hash values aim at equal distribution of RDD partitions between executors, but provide no guarantee on achieving data locality desired by our solvers.

\begin{figure}
\centering
\includegraphics[scale=1.15]{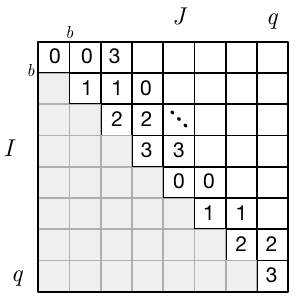}
\caption{Data layout induced by our multi-diagonal partitioner. Blocks with the same index are assigned to the same RDD partition.}\label{fig:partfunc}
\end{figure}

The second partitioner is our multi-diagonal partitioner (MD). The assignment of matrix blocks to partitions induced by this approach is outlined in Figure~\ref{fig:partfunc}. The idea behind MD partitioner is to explicitly place blocks that are in the same column- and row-block into separate partitions, while maintaining equal block distribution between the partitions. The strategy is especially critical to Blocked In-Memory and Blocked Collect/Broadcast methods, as it enables us to avoid bottlenecks in Phase 2 of these algorithms. At this point we should note that MD partitioner may seem counter-intuitive from the perspective of matrix algorithms based on 2D decomposition. Typically, such algorithms involve some form of grid-based partitioning to optimize communication patterns. However, as we mentioned earlier, in Spark we have limited control over which executors handle which partitions. Therefore, the best strategy is to use partitioning which will give scheduling flexibility to the Spark~runtime.

In Table~\ref{blockmethodstable0} we quantify the effect of the block size and the partitioner on the performance of each solver. Here several observations stand out. First, Repeated Squaring and 2D Floyd-Warshall both are infeasible on large problems, with projected execution times going into days. The poor performance is especially pronounced for 2D Floyd-Warshall. Even though from the computational point of view the method is asymptotically comparable to the blocked methods, it suffers from unfavorable balance between computations and data movement -- each iteration involves $O(b^2)$ operations, the number of iterations is linear in the problem size. Repeated Squaring on the other hand shows very good performance in single iteration, but this is insufficient to compensate $\log(n)$ higher number of iterations (compared to the blocked methods).

The second observation is that the effect of partitioner depends on the block size, and on factor $B$. When the block size is small, the difference between PH and MD partitioners becomes negligible. This is because partitions become fairly balanced just by chance, as we have many blocks to distribute, and even when partitions are slightly unbalanced, the difference in runtime becomes marginal owing to the small block size. Conversely, for the large block size the impact of the partitioner on runtime becomes critical. Figure~\ref{fig:blockandpart} (bottom plot) explains this effect in case of blocked methods. The PH partitioner consistently fails to evenly distribute blocks to RDDs, which is caused by suboptimal choice of the hashing function (inspection of Spark code reveals that it uses XOR based mixing of elements of the tuple, which in case of upper-triangular matrix leads to many collisions). The resulting skew in hash distribution translates directly into poor runtimes (Figure~\ref{fig:blockandpart} top plot). This is especially pronounced for the Blocked In-Memory method when $B=1$. In this case we have to deal with very large and unbalanced partitions with no room for load balancing by dynamic scheduling. Additionally, we run multiple repartitioning steps with their shuffling operations, which ultimately result in unbalanced spills to local storage. We note that because Blocked Collect/Broadcast mitigates shuffling, it performs significantly better. 

The last observation is with respect to the over-decomposition factor, $B$. From our results it is clear that the Spark guideline of having $B > 1$ is imperative. However, since in our solvers the total number of blocks to distribute is a function of the block size, for larger $p$ we may easily end up with $B=1$. In such situations, it may be advantageous to scale down $p$ to increase $B$ without decreasing block size $b$. This strategy however ultimately limits the scalability of the method. In our tests, most of the time we were able to maintain $B=2$.

\subsection{Scalability Tests}\label{sec:scalability}

We now turn our attention to scalability of the solvers. We focus only on Blocked In-Memory and Blocked Collect/Broadcast, since the other two methods cannot be expected to scale to larger problems (see Table~\ref{blockmethodstable0}). We perform weak scaling analysis, as it can be considered a typical use scenario for Spark (i.e., we increase cluster size to solve larger problem instances).

To assess scalability we measure operations per second (ops), which we express as $\displaystyle\frac{n^3}{T_p}$, where $T_p$ is the time taken to solve the APSP problem of size $n$ using $p$ cores. Considering that the key computational building blocks used by our solvers have complexity~$O(n^3)$, the measure is a good proxy to assess the performance. For convenience we are reporting giga-ops (Gops) normalized with respect to $p$ (Gops/core). Our reference point is $T_1$ -- the time taken for a proportional problem on a single core using efficient sequential Floyd-Warshall as implemented in SciPy and tested in Section 5.2.

\begin{figure}
\centering
\includegraphics[scale=1.05]{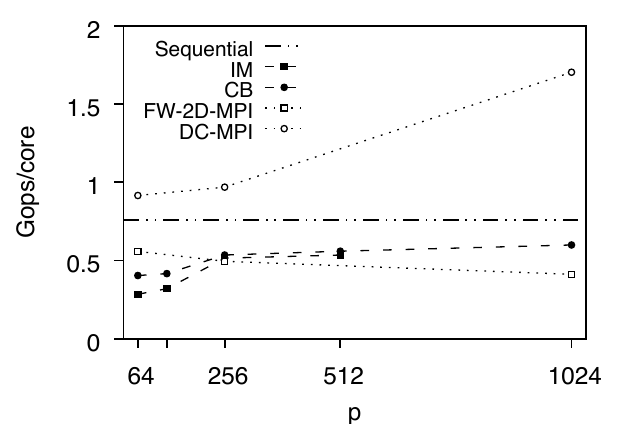}
\caption{Weak scaling of blocked methods.}\label{fig:weakops}
\end{figure}

Results of the experiment are summarized in Table~\ref{weakscalingtable} and Figure~\ref{fig:weakops}. Here we maintain $\displaystyle\frac{n}{p} = 256$ to ensure that the largest resulting problem (i.e., $n=262144$) is feasible given the hardware resources we use in the tests. Moreover, for each method and every problem size, we use our multi-diagonal partitioner, and we select the optimal block size, $b$, following the arguments given in the previous sections (we report block size in Table~\ref{weakscalingtable}). Finally, for $n=256$ we record $T_1 = 0.022$s, which translates into $0.762$Gops.

Our results show that as expected Blocked Collect/Broadcast outperforms Blocked In-Memory. Moreover, for the largest problem size ($p=1024$), Blocked In-Memory runs out of space in the local storage, and is unable to finish processing. Both methods show rather stable scaling that saturates around $p=256$. For $p<256$ performance is slightly degraded, which we attribute to Spark scheduler: for the small problem size, the number of RDD partitions per core is one, and that limits optimal resources utilization (note however, that choosing different block size does not improve performance). For $p=1024$, the Blocked Collect/Broadcast solver achieves 78\% Gops/core with respect to the sequential solver, which we consider quite good result taking into account all Spark-added overheads. We should keep in mind though that Blocked Collect/Broadcast is not fault-tolerant.

\begin{table}
\caption{Weak scaling of blocked methods.}\label{weakscalingtable}
\centering
\renewcommand{\arraystretch}{1.2}
{\small
\begin{tabularx}{\columnwidth}{lXXXXX}
\toprule
Method~~~~~/~~~~~$p$ & 64 & 128 & 256 & 512 & 1024 \\
\midrule
Blocked-IM & 4m2s & 14m20s & 35m33s & 2h17m & --\\
~$b$ & 1024 & 1024 & 1536 & 2048 & --\\
\cmidrule{1-1}
Blocked-CB & 2m50s & 11m0s & 34m16s & 2h11m & 8h9m\\
~$b$ & 1024 & 1280 & 1536 & 2048 & 2560\\
\cmidrule{1-1}
FW-2D-GbE & 2m3s & -- & 37m2s & -- & 11h51m\\
DC-GbE & 1m15s & -- & 18m54s & -- & 2h52m\\



\bottomrule
\end{tabularx}}

\smallskip
{\scriptsize $\frac{n}{p}=256, T_1=0.022$s}
\end{table}%

\subsection{Comparison with MPI-based Solvers}

In the last set of tests we contrast performance of our solvers with the approaches based on MPI. While this may seem like ``comparing apples with oranges,'' the exercise helps to provide a reference point to judge Spark scaling. Since MPI-based optimized APSP solvers have been around for some time, such a reference point is of value especially to the Machine Learning community that is predominantly using Spark.

To make the comparison, we use two MPI-based APSP solvers. In the first solver, FW-2D-GbE, we implement the standard parallel Floyd-Warshall algorithm based on 2D block decomposition~\cite{Grama2003} (described in Section~4.3). The method is relatively straightforward to implement, however, as usual for MPI, the resulting code is objectively more verbose and complex than its Spark counterpart. The second method, abbreviated as DC-GbE, is highly optimized divide-and-conquer solver by Solomonik et~al.~\cite{Solomonik2013}, available from \url{https://github.com/solomonik/APSP}. The solver has been demonstrated to scale extremely well to very large parallel machines, and is the state-of-the-art HPC solution. From the code complexity perspective, this solution is the most elaborate. Both solvers are written in C++, and we compile them with Intel~MPI~2018 and g++~7.8 for execution. Both are run on the same cluster as Spark. Finally, since both MPI solvers assume that processors are organized into a square grid, in our tests we use $p \in [64,256,1024]$.

From Table~\ref{weakscalingtable} and Figure~\ref{fig:weakops} we can see that Spark-based solvers outperform naive MPI-based solution for larger problem sizes. This poor performance of MPI-based Floyd-Warshall is explained by communication overheads, specifically latency, that grow with $log(p)$ (due to broadcast) and $n$ (due to $n$ iterative steps). However, we should not forget that for the same reasons the method is completely infeasible in Spark. On the other hand, the optimized DC solver significantly outperforms (over $2.8\times$ on $p=1024$ cores) all other solutions. While this result is not surprising, it clearly highlights that a good runtime system is required but not sufficient to obtain good scalability -- improving data locality and reducing communication, which are the key optimizations used by the DC solver and by our Blocked methods, are essential for scalability.







\section{Conclusion}\label{sec:conclusion}


In this paper, we proposed and analyzed four different implementations of APSP solver using Apache Spark. We demonstrated that by separating core computational functionality from the data and tasks distribution problem, we can implement APSP in pySpark in just few compact lines of code (including usual boilerplate). Our Blocked-CB method is able to handle large graphs in acceptable time limits, but it requires compensating missing Spark functionality (e.g., broadcasts from arbitrary source) with solutions outside of the Spark API. Consequently, the method depends on persistent storage for data broadcasting and thus is not fault-tolerant.

\subsection{Takeaway Points}

To conclude our presentation, we wish to offer several takeaway points regarding design and implementation of APSP in Apache Spark. First, while the Spark API makes it look easy to express a solver and forget about volume of the data this solver must handle, technical nuances make it equally easy to deliver an inefficient solution. For example, disregarding how RDD lineages evolve, and how RDD unions are realized, leads to over-partitioning that in turn overloads Spark scheduler (we mention this issue in Section~5.2). Hence, programmer must be constantly aware of how different Spark transformations and actions are realized to avoid or mitigate unnecessary overheads. Second, data partitioning and communication (or data movement) are as critical as in any HPC-oriented platform. For example, carefully choosing block size $b$ and using a new multi-diagonal partitioning scheme (Section~5.3) were critical to scale our solvers. Thus, programmer should not depend on {\it default} options provided by Spark, which most likely will be suboptimal. Finally, high-level problem decomposition that isolates computationally heavy elements from those responsible for computations coordination is essential. Without that decomposition (given in Table~1) we would not be able to achieve as compact Spark solvers as we presented. Furthermore, this decomposition enables us now to investigate hybrid models combining Spark with, e.g., GPGPUs.



\section{Acknowledgments}

The authors would like to acknowledge support provided by the Center for Computational Research at the University~at~Buffalo. This work has been supported by the National Science Foundation under grant OAC-1910539.



%
\bibliographystyle{ACM-Reference-Format}
\bibliography{sample-base}

\end{document}